\documentclass[12pt]{article}
\usepackage{latexsym}
\usepackage{epsfig}
\usepackage{amsmath}
\usepackage{graphicx}
\usepackage{amsfonts}
\usepackage{amssymb}
\newtheorem{Proposition}{Proposition}
\newtheorem{Theorem}{Theorem}
\newtheorem{Lemma}{Lemma}
\begin{document}

\title{The Dynamics of the Linear Random Farmer Model}
\author{Rui Carvalho\thanks{I\ would like to acknowledge long discussions with Rui
Vilela Mendes and suggestions by an anonymous referee.}\\{\small Laborat\'{o}rio de Mecatr\'{o}nica, Instituto Superior T\'{e}cnico}\\{\small Av. Rovisco Pais, 1049-001 Lisboa Portugal}\\{\small and }\\{\small The Bartlett School of Graduate Studies, University College London}\\{\small Gower Street, London WC1E 6BT UK}\\rui.carvalho@ucl.ac.uk}
\date{{\small 22th of June 2001}}
\maketitle
\begin{abstract}
\textsf{On the framework of the Linear Farmer's Model, we approach the
indeterminacy of agents' behaviour by associating with each agent an
unconditional probability }$\mathsf{p}^{(i)}\mathsf{\ }$\textsf{for her to be
active at each time step. }

\textsf{We show that Pareto tailed returns can appear even if value investors
are the only strategies on the market and give a procedure for the
determination of the tail exponent. }

\textsf{Numerical results indicate that the returns' distribution is heavy
tailed and volatility is clustered if trading occurs at the zero Lyapunov
(critical) point. }
\end{abstract}

\section{\bigskip Introduction}

Share prices and foreign exchange rates (or their logarithms) obey three
stylized facts \cite{Luxinpress}: price variations are uncorrelated (thus, one
is unable to reject the hypothesis that financial prices follow a random walk
or a martingale); on the other hand, the amplitude of price variation
(volatility) is non-homogeneous and its correlations are very
long-ranged\label{intropage}, such that large (price) changes are followed by
large changes --of either sign-- and small changes tend to be followed by
small changes (these periods of quiescence and turbulence are known in the
finance literature as volatility clusters); finally, the returns unconditional
distributions are 'fat tailed' (they decay slower than a Gaussian).

Recently, there have been a number of related studies investigating the
dynamical behaviour in heterogeneous belief models \cite{Beja80, Day90,
Chiarella92, Lux95, Cabrales96, Sethi96, Lux98}. In these studies, two typical
classes of agents are fundamentalists, expecting prices to return to their
'fundamental value', and chartists or technical analysts extrapolating
patterns, such as trends, in past prices. To our knowledge, this research
agenda can be traced back to the work of Beja and Goldman \cite{Beja80}. They
studied an ''out of equilibrium'' model with linear strategies and concluded
that the system is stable only when fundamental strategies are largely
dominating. Day and Huang \cite{Day90} studied a related model where the
market maker was introduced as a participant in the dynamics and nonlinear
investment rules were admitted. They discussed how this could lead to chaotic
price series. These works were purely deterministic.

Lux proposed a disequilibrium model that allows stochastic transitions from
one strategy to another, in accordance with their respective performance
\cite{Lux98}. He showed that the high peaks and fat tails property can be
derived from the endogenous dynamics itself. Lux and Marchesi \cite{Lux99}
extended this work further by introducing a model such that the market
dynamics transforms exogenous noise (news) into a fat tailed unconditional
distribution for the returns with clustered volatility.

Gaunersdorfer and Hommes observed that there are fundamentally two well-known
concepts in nonlinear dynamics which appear to be appropriate for a
description of volatility clustering: intermittency and multistability
\cite{Gaunersdorfer2000}. Although the authors only mention Pomeau-Manneville
intermittency, we will be interested in one type of intermittency
characteristic of random dynamical systems\footnote{A random map can be
defined in the following way. Denoting with $\mathbf{x}(t)$ the state of the
system at discrete time $t$, the evolution law is given by
\[
\mathbf{x}\left(  t+1\right)  =\mathbf{f}\left(  \mathbf{x}\left(  t\right)
,J\left(  t\right)  \right)
\]
where $J\left(  t\right)  $ is a random variable. For an introduction to
random dynamical systems, see \cite{Boffetta2001, Arnold98}.}: \emph{on-off
intermittency}. On-off intermittency in dynamical systems is triggered by the
repeated variation of one dynamical variable through the bifurcation point of
another dynamical variable. The first variable acts as a time-dependent
parameter, while the response of the second variable comprises the
intermittent signal\cite{Heagy94}. This mechanism can be easily recreated in
dynamical systems that are skew products\cite{Platt93}, including random
dynamical systems. The importance of on-off intermittency is based on the
growing accumulating evidence that chaotic transitions through on-off
intermittency may be one of the routes to chaos of random dynamical systems as
period doubling, intermittency, quasiperiodicity, and crises are those of
deterministic dynamical systems\cite{Rim2000}.

Recently, there has been a burst of work in heterogeneous models providing
possible explanations to volatility clustering in financial markets. Iori
\cite{Giulia2001} proposed a model where large fluctuations in returns arise
purely from communication and imitation among traders. The key element in the
model is the introduction of a trade friction (representing transactions
costs) which, by responding to price movements, creates a feedback mechanism
on future trading and generates volatility clustering. Through massive
simulations of autonomous agents and the use of evolutionary (genetic)
algorithms, both Arifovic and Gen\c{c}ay \cite{Arifovic2000} and Arthur et al.
\cite{Arthur97} use feedback mechanisms with an adaptative feature to account
for the presence of volatility clustering in financial time series.
Gaunersdorfer, Hommes and Wagener\cite{Gaunersdorfer2000a} conjecture that
intermittency and coexistence of attractors, are also relevant to other
computationally oriented nonlinear evolutionary multi-agent systems such as
the Santa Fe Artificial stock market \cite{Arthur97}. Arguably, the best
accomplished result of these evolutionary models has been an endogenous
explanation for volatility clustering, in contrast with the classical approach
in empirical finance where volatility clustering is modeled as an exogenous
phenomenon by a statistical model\footnote{The main problem with GARCH models
is that they are ad hoc models and do not relate the variables to an economic
context.} or one of its extensions.

J. Doyne Farmer has developed a microscopic market model based on a
non-equilibrium price formation rule and a coevolving ecology of trading
strategies \cite{Farmer98}. The theory developed by Farmer is based on a
deduction of the market impact function and a first order (linear) Taylor
series expansion of the strategies (value investors and trend followers).
Farmer showed numerically that state dependent threshold (nonlinear)
deterministic strategies and their evolution could be possible mechanisms for
an explanation of the stylized facts observed in financial time series. Farmer
and Joshi rewrited the former analysis, omitting the evolutionary approach,
but grounding the arguments in a more solid economic literature review
\cite{Farmer2001}. The authors observed clustered volatility, but stated that
it is not clear if the resulting volatility correlations are strong enough to
match those observed in real data. Bouchaud and Cont modeled the price
dynamics in a similar way \newline through a ''Langevin model''
\cite{Bouchaud98}.

The main goal of this paper is to address the explanation of the stylized
facts of finance on the framework of a simple affine random dynamical system,
e.g. Farmer's model with linear strategies under the added hypothesis that
traders are stochastic. Stochastic traders have a (fixed) unconditional
probability to be active at each time step. As we shall see, Pareto tailed
returns can appear even if value investors are the only strategies on the
market. Further, we analyze the model with interacting trend followers and
value investors. If the largest Lyapunov exponent of the (linear) market
impact function is zero\footnote{A positive Lyapunov exponent is the
operational definition of chaos and is the primary dynamical invariant used to
characterize a chaotic process. A zero Lyapunov exponent corresponds to the
marginal case between exponential growth and exponential decay. The parameter
value for which the system has a zero Lyapunov exponent is a non-hyperbolic or
critical point.}, the system will display on-off intermittency and the time
series of the returns will display 'realistic' \cite{Gopikrishnan98} tails in
the range $2\sim4$. The model generates clustered volatility, but, as we shall
show, the volatility autocorrelation function decays exponentially and not as
a power-law. We also observe that the price series displays extremely wild
oscillations not in agreement with a model where price and fundamental value
cointegrate. The extent up to which this is relevant may be questionable, as
stock prices and dividends series do not seem to cointegrate universally
\cite{Sarno99}.

The introduction of an unconditional probability for the agents' activity is
justified on the following possible sources of uncertainty: inherent
unpredictability of agents' behaviour (random preferences) and lack of
knowledge on the part of the econometrician (random characteristics)
\cite{Brock2000}. Traditionally, critical states were thought to be associated
with certain ''critical'' parameter values (such as the temperature), that
would only occur if they were ''tuned'' to be at the critical value in a
laboratory experiment. Recently, it has been argued that the critical state
can actually be an attractor for the dynamical system, toward which the system
''naturally'' evolves \cite{Bak88}. Bak, Chen, Scheinkman and Woodford
presented a simple model of an economic system in this direction \cite{Bak93}.
In this paper we will assume that the\emph{\ }critical point is the asymptotic
state of a self-organizing dynamics, but we will not study the later.

Therefore, we define the random linear Farmer's model as the model presented
on \cite{Farmer98}\ where agents $i=1\ldots N$ \ follow linear position based
value investor or trend follower strategies, $\omega^{(i)}$, which are active
with probability (\emph{w.p.}) $p^{(i)}$. Nevertheless, it is outside of the
scope of this paper to consider dynamical effects caused by evolution or
capital reinvestment. Indeed, traders will have constant capital and strategies.

The remainder of this paper is organized as follows. Section 1 introduces
Farmer's model and our modification. Section 2 analyses rigorous conditions
for the returns distribution to be 'fat tailed' when only value investor
strategies are present. Section 3 shows that although the combination of trend
followers and value investors generates a random walk in mispricing, the
return's statistics on the (non-hyperbolic) critical point displays fat tails
and volatility clustering. Section 4 concludes.

\section{\bigskip The Model}

\subsection{The Market Impact Function}

\bigskip Consider a stock market with $N$ agents, labeled by an integer $1\leq
i\leq N$, trading on a single asset (measured in units of shares), whose price
at time $t$ will be denoted $P_{t}$. Let $p_{t}=\log P_{t}$ be the log price,
$r_{t+1}=p_{t+1}-p_{t}$ be the one-period return and $\omega^{(i)}%
(p_{t},p_{t-1},\ldots,I_{t})$ \ be the order size for trader $i$ at time step
$t,$ where $I_{t}$ represents any additional external information.

The trade protocol in Farmer's model involves a two-step process. On the first
step, the asset's true value ($\nu$) is realized and each trader posts their
orders based on this fundamental value, the current ($p_{t}$) and past
($p_{t-1},\ldots$) prices. On the second stage, the market maker observes the
aggregate net order flow%

\[
\omega=\sum_{i=1}^{N}\omega^{(i)}%
\]
and sets a single price ($p_{t+1}$) according to the \emph{market impact
function} (also called the \emph{price impact function}). Orders are then
filled at the new price, $p_{t+1}$.

Farmer deduces the market impact function to be \cite{Farmer98}
\begin{equation}
p_{t+1}=p_{t}+\frac{1}{\lambda}\sum_{i=1}^{N}\omega_{t+1}^{(i)}(p_{t}%
,p_{t-1},\ldots,I_{t})+\xi_{t+1} \label{market impact equation}%
\end{equation}
where $i$ indexes the trader, $\lambda$ is the market maker liquidity and
$\xi_{t}$ accounts for exogenous influences on the dynamics\footnote{On the
following analysis, we will assume $\xi_{t}\equiv0$.}. \ Under equation
(\ref{market impact equation}), the price is manipulated in the direction of
the net of incoming orders \cite{Farmer98}.

\subsection{The Market Metaphor}

The dynamics of (\ref{market impact equation}) depends on the collection of
strategies, which Farmer divides into two classes \cite{Farmer98}. Traders can
be \emph{fundamentalists} or \emph{technical analysts}. Fundamentalists
believe that the price of an asset is determined solely by its efficient
market hypothesis (EMH) fundamental value. Technical analysts, believe that
asset prices are not completely determined by fundamentals, but that they may
be predicted by simple technical trading rules, extrapolation of trends and
other patterns observed in past prices \cite{Hommes98}.

The model builds upon a vision of the market based on interaction between
micro and macro dynamics. The former is the dynamics of individual agents, who
invest according to \emph{fundamental} or \emph{technical} strategies. The
later is the dynamics of fundamental value, a random walk for the scope of
this text, and prices. Traders do not communicate with each other; instead
they interact with the market maker 'through' equation (\ref{market impact
equation}) and base their decisions upon present and past values of the price
or fundamental value. Therefore, traders' decisions (the finegrained
variables) are influenced by prices and fundamental values (the coarse grained
variables), and the following price is determined by traders' investments. The
market is born out of this feedback loop between finegrained and coarsegrained variables.

\subsection{The Strategies}

A first class of agents, the value investors, have pure \emph{position based
value strategies} which can be approximated by\footnote{Farmer \cite{Farmer98}
ellaborates an analysis of order-based and position-based value investors. He
concludes that order-based strategies (the equivalent of the $\beta
$\emph{-investors} for Day and Huang\cite{Day90}) are unrealistic as the
trader can accumulate unbounded inventory.}%

\begin{equation}
\omega_{n+1}^{(i)}\simeq-c^{(i)}\Delta m_{n}=-c^{(i)}(r_{n}-\eta_{n})
\label{vi}%
\end{equation}
where $m_{n}=p_{n}-\nu_{n}$ is the difference between (log) price and (log)
fundamental value, $\Delta m_{n}=m_{n}-m_{n-1}$ ($\Delta$ is the difference
operator), $r_{n}=p_{n}-p_{n-1}$ is the one-period return and $\eta_{n}%
=\Delta\nu_{n}$ is the random increment to value.

The second class of agents attempt to identify price trends as a source of
information, instead of focusing on fundamentals. Farmer calls these the
\emph{trend following strategies}. An example of a \emph{position based}
\emph{trend following strategy} on timescale $\theta$ is
\begin{equation}
\omega_{n+1}^{(j)}=c^{(j)}(r_{n}-r_{n-\theta}) \label{tf}%
\end{equation}

We consider a system with $N_{v}$ agents holding position based value
strategies and $N_{t}$ agents with position based trend following strategies,
so that $N=N_{v}+N_{t}$ is the total number of agents. Value investors are
indexed $1,\ldots,N_{v}$ and trend followers are indexed $N_{v}+1,\ldots,N$.

For the sake of the theoretical analysis, we assume that all parameters are
constant. In this way, we can concentrate on the consideration of market
behaviour in its relation to pure fundamental/speculative forces and price adjustments.

Replacing (\ref{vi}) and (\ref{tf}) on (\ref{market impact equation}),
defining the sums of the agents normalized capital for each class of
strategies as $\alpha^{vi}=\frac{1}{\lambda}\sum_{i=1}^{N_{v}}c^{(i)}$ for
value investors and $\alpha^{tf}=\frac{1}{\lambda}\sum_{j=N_{v}+1}^{N}c^{(j)}
$ for trend followers\footnote{We will drop the subscripts $vi$ and $tf$,
whenever they are superfulous.}, and the difference between the value
investors' and trend followers' total normalized capital as $\Delta
\alpha=\alpha^{vi}-\alpha^{tf}$, yields:%

\begin{equation}
\left\{
\begin{array}
[c]{l}%
r_{n+1}=-\Delta\alpha\,r_{n}-\alpha\,^{tf}r_{n-\theta}+\alpha^{vi}\,\eta_{n}\\
p_{n+1}=p_{n}+r_{n+1}%
\end{array}
\right.  \label{market impact RC}%
\end{equation}

Farmer allows the trend follower's $\theta$ to be agent dependent, that is
$\theta\equiv\theta^{(i)}$. As we shall observe in Section $4$, this affects
equation (\ref{market impact RC}).

To the above framework, we add to each agent $i$ a probability for her to
trade (either buy or sell) at each time step, $p^{(i)}$. Therefore, the market
has a non-constant (in time) number of participants. It will be useful to
define the following
\begin{equation}%
\begin{array}
[c]{lll}%
\delta_{\theta^{(i)},n} & =1 & \text{\textrm{if trend follower} }%
i\mathrm{\ }\text{\textrm{has time lag} }\theta^{(i)}\mathrm{\ }%
\text{\textrm{and}}\\
&  & \text{\textrm{is active at time step} }n\mathrm{\ }\text{\textrm{and}
}0\mathrm{\ }\text{\textrm{otherwise;}}\\
\delta_{i,n} & =1 & \mathrm{\ }\text{\textrm{if trader} }i\mathrm{\ }%
\text{\textrm{is active at time step} }n\mathrm{\ }\text{\textrm{and}
}0\mathrm{\ }\text{\textrm{otherwise}}%
\end{array}
\label{delta}%
\end{equation}

\subsection{Value Investors Only\label{sectionvi}}

On this section we assume that the only strategy is value investing. Under the
assumptions that the traders' normalized capital and trading probability are
constants, we show that the returns' unconditional distribution can display
Pareto tails and determine an expression for the tail exponent. A simple
physical explanation is given for this effect. We refer the reader to Appendix
B for proof of the theorems.

Define the total value investors' and trend followers normalized capital
respectively as $\alpha_{n}^{vi}=\sum_{i=1}^{N_{v}}\alpha^{(i)}\delta_{i,n}$
and $\alpha_{n}^{tf}=\sum_{i=N_{v}+1}^{N}\alpha^{(i)}\delta_{i,n}$.

Consider $N=N_{v}$, that is only value investors are present, and define
$\alpha_{n}\equiv\alpha_{n}^{vi}$. Then equation (\ref{market impact RC})
becomes
\begin{equation}
r_{n+1}=-\alpha_{n}r_{n}+\alpha_{n}\eta_{n} \label{eq vi}%
\end{equation}

Consider that the agents' normalized capital and trading probability are
constants, respectively $\alpha_{n}^{(i)}=\alpha$ and $p^{(i)}=p$. Then
$\alpha_{n}$ has a binomial distribution,
\[
\alpha_{n}=\left\{
\begin{array}
[c]{ll}%
0 & \emph{w.p.}\text{ }(1-p)^{N}\\
\alpha & \emph{w.p.}\text{ }p(1-p)^{N-1}N\\
\vdots & \vdots\\
(N-1)\alpha & \emph{w.p.}\text{ }p^{N-1}(1-p)\binom{N}{N-1}\\
N\alpha & \emph{w.p.}\text{ }p^{N}%
\end{array}
\right.
\]

Define the intervals
\begin{equation}
I_{k,k+1}=\left]  \frac{1}{\sqrt[k+1]{M^{(k+1)}(0)}},\frac{1}{\sqrt[k]%
{M^{k}(0)}}\right[
\end{equation}
Where $M^{(k)}(0)$ is the $k^{th}$ moment of the distribution of $\alpha_{n}
$. We will show that if one chooses $\left\{  N,k\right\}  \in\mathbb{N}^{2}$,
$p\in\left]  0,1\right[  $ and $\alpha\in I_{k,k+1}$, then the tails of the
limiting distribution of $r_{n}$ are asymptotic to a power law with exponent
$-\mu$ for a certain $\mu\in\left[  k,k+1\right]  $ and $p$ small enough.

The proof will follow three steps. First we show (Lemma \ref{lemma0}) that the
$I_{k,k+1}$ are ordered intervals. Second we show (Lemma \ref{lemma1}) that
for $\alpha\in I_{k,k+1}$and $k\in\mathbb{N}$, $r_{n}$ converges in
distribution to a unique limiting distribution. Finally, we state Kesten's
Theorem (Theorem \ref{theorem Kesten}) which gives conditions for the presence
of fat tails and permits the determination of the tail exponent on
one-dimensional affine random dynamical systems. Although we do not determine
a simple closed form expression for the tail exponent, we show that it can be
bounded by two adjacent integers (Proposition \ref{proposition}).

\begin{Lemma}
\label{lemma0}$\exists p_{0}\in]0,1[\ \forall p<p_{0}$%
\begin{equation}
\frac{1}{\sqrt[k+1]{M^{(k+1)}(0)}}<\frac{1}{\sqrt[k]{M^{(k)}(0)}}
\label{Square Root Moment}%
\end{equation}
\end{Lemma}

\begin{Lemma}
\label{lemma1}Choose $\left\{  N,k\right\}  \in\mathbb{N}^{2}$, $p\in\left]
0,1\right[  $ and $\alpha\in I_{k,k+1}$. Then $r_{n}$ converges in
distribution to a unique limiting distribution.
\end{Lemma}

\medskip

Equation (\ref{eq vi}) is well know to present Pareto tails. In fact, Kesten
\cite{Kesten73} states the following

\begin{Theorem}
[Kesten (simplified)]\label{theorem Kesten}\bigskip Consider a stochastic
difference equation:
\[
x_{t+1}=m_{t}\ x_{t-1}+q_{t}\qquad t=1,2,\ldots
\]
where the pairs $\left\{  m_{t},q_{t}\right\}  $ are $i.i.d.$ real valued
random variables. \newline If $x_{t}$ converges in distribution to a unique
limiting distribution, $q_{t}/\left(  1-m_{t}\right)  $ is non-degenerate and
there exists some $\mu>0 $ such that: \newline i) $0<\left\langle \left|
q_{t}\right|  ^{\mu}\right\rangle <\infty$ \newline ii)$\left\langle \left|
m_{t}\right|  ^{\mu}\right\rangle =1$ \newline iii) $\left\langle \left|
m_{t}\right|  ^{\mu}\log^{+}\left|  m_{t}\right|  \right\rangle <\infty$
\newline then the tails of the limiting distribution are asymptotic to a power
law, \emph{i.e.} they obey a law of the type
\begin{equation}
Prob\left(  \left|  x_{t}\right|  >y\right)  \approx c\cdot y^{-\mu}
\label{power law}%
\end{equation}
\end{Theorem}

The derivation of (\ref{power law}) uses results from renewal theory of large
positive excursions of a random walk biased towards $-\infty$ (see
\cite{Feller71}, sections $VI6-8$, $XI1$, $XI6$, $XII4b$ and $XII5$ for an
outline of the proof when $q_{t}$ is positive) .

\begin{Proposition}
\label{proposition}Choose $\left\{  N,k\right\}  \in\mathbb{N}^{2}$,
$p\in\left]  0,1\right[  $ and $\alpha\in I_{k,k+1}$. Then the tails of the
limiting distribution of $r_{n}$ are asymptotic to a power law with exponent
$-\mu$ for a certain $\mu\in\left[  k,k+1\right]  $ and $p$ small enough,
\emph{i.e.} they obey a law of the type $P(\left|  r_{n}\right|  >x)\approx
c\cdot x^{-\mu}$\quad$\exists\mu\in\left[  k,k+1\right]  $.
\end{Proposition}

\noindent\medskip

Intuitively, heavy tails appear on the system as the expected value of the
traders capital is less than one ((\ref{eq vi}) has a negative Lyapunov
exponent), but it attains very high values as a massive number of traders
simultaneously enter the market (this happens intermittently), causing
explosions in price.

Numerical calculations show that for $N=10^{3}$, $p=9.5\cdot10^{-3}$ and
$\alpha=0.1$, $\left\langle \alpha_{n}\right\rangle =.95$, $\left\langle
\left(  \alpha_{n}\right)  ^{2}\right\rangle =0.996598$ and $\left\langle
\left(  \alpha_{n}\right)  ^{3}\right\rangle =1.13478$. Thus, one should
expect a $\mu$ very close to $2$.

\begin{figure}
[ptb]
\begin{center}
\includegraphics[
height=3.9141in,
width=4.7469in
]%
{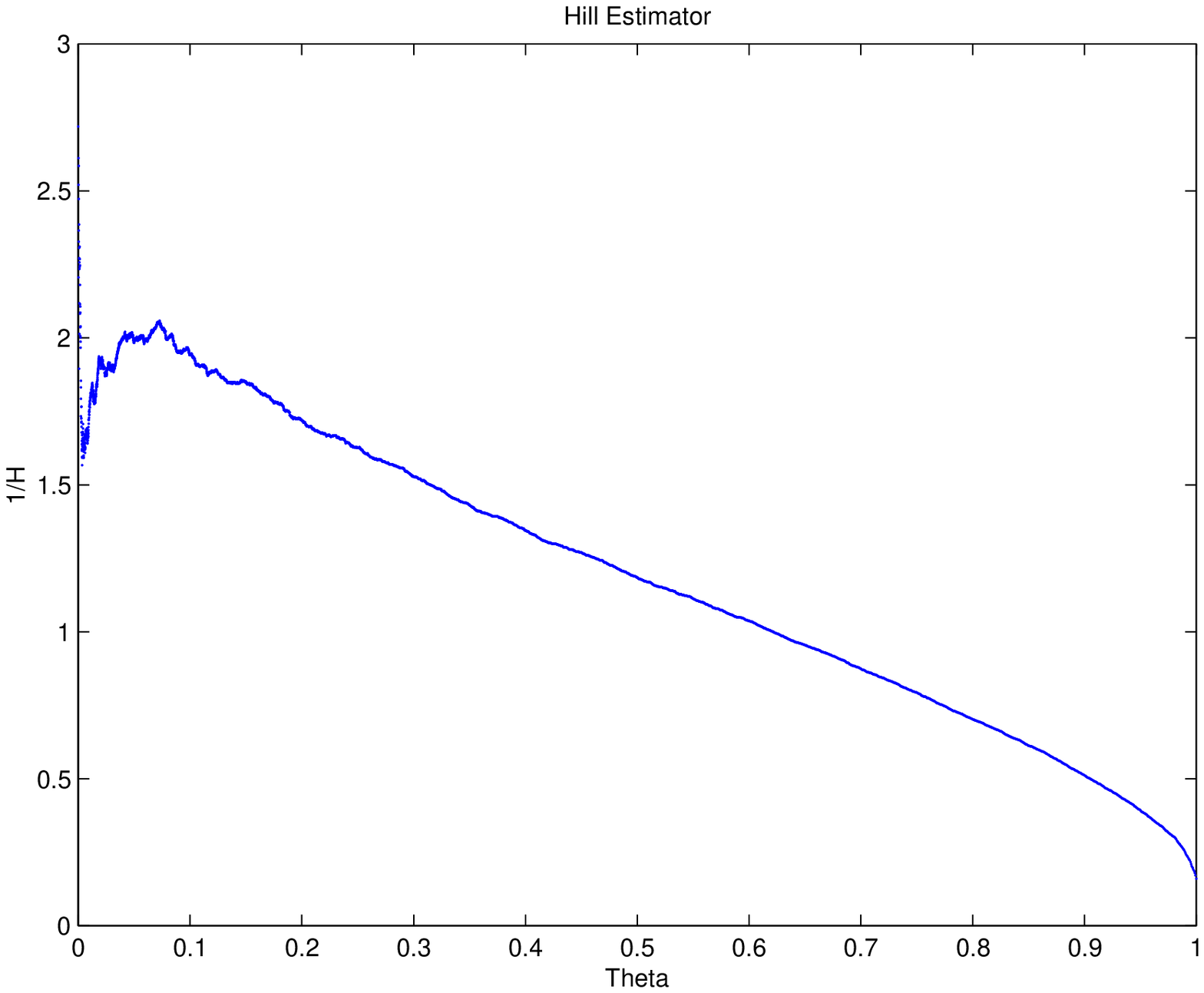}%
\caption{Hill estimator for the tails of $r_{n}$ when $N=10^{3}$,
$p=9.5\cdot10^{-3}$ and $\alpha=0.1$.}%
\label{HillFig}%
\end{center}
\end{figure}

Figure \ref{HillFig} is the modified Hill plot (see Apendix C) for $r_{n}$
which numerically confirms these calculations.

\smallskip

\subsection{Trend Followers and Value Investors Trading on the Critical Point}

To better understand (\ref{market impact RC}), we notice that
\[
r_{n+1}=\triangle m_{n+1}+\eta_{n+1}%
\]
so that the dynamics of the mispricing is
\begin{equation}
\triangle m_{n+1}=-\triangle\alpha\ \triangle m_{n}-\alpha^{tf}\ \triangle
m_{n-\theta}-\eta_{n+1}+\alpha^{_{tf}}\ \eta_{n}-\alpha^{_{tf}}\eta_{n-\theta}
\label{mispricing dynamics}%
\end{equation}
Equation (\ref{mispricing dynamics}) \ is a discrete dynamical system with
delay structure. By making the substitution%

\begin{equation}
(X_{n}^{1},\ldots,X_{n}^{\theta},X_{n}^{\theta+1})\equiv(\triangle
m_{n-\theta},\ldots,\triangle m_{n-1},\triangle m_{n}) \label{X replaces z}%
\end{equation}
the dynamics of $\triangle m_{n}$ becomes
\begin{equation}
\mathbf{X}_{n+1}=\mathbf{J}_{m}\mathbf{\ X}_{n}+\mathbf{\eta}_{n}
\label{Eq LAG}%
\end{equation}
where $\mathbf{J}_{m}$ is a $(\theta+1)\times(\theta+1)$ companion matrix,
\begin{equation}
\mathbf{J}_{m}=\left[
\begin{array}
[c]{ccccc}%
0 & 1 &  &  & \\
&  & 1 &  & \\
&  &  & \vdots & \\
&  &  &  & 1\\
-\alpha^{tf} & 0 & \ldots & 0 & -\Delta\alpha
\end{array}
\right]  \label{Jm}%
\end{equation}
and%

\begin{equation}
\mathbf{\eta}_{n,\theta}=\left[
\begin{array}
[c]{l}%
0\\
\vdots\\
-\eta_{n+1}+\alpha^{_{tf}}\ \eta_{n}-\alpha^{_{tf}}\ \eta_{n-\theta}%
\end{array}
\right]  \label{greek1}%
\end{equation}

If the system has $N_{tf}$ trend followers, equation (\ref{Jm}) is time
dependent, $\mathbf{J}_{m}\mathbf{\equiv J}_{m}(n)$.

The sum of the capital of active trend followers with time lag $\theta^{(i)} $
at time step $n$ is
\[
\alpha_{n}^{tf}(\theta^{(i)})=\sum_{k=1}^{N_{t}}\alpha_{n}^{(k)}\delta
_{\theta^{(i)},n}%
\]
and the difference in capital between (active) value investors and trend
followers total capital at time step $n$ is
\[
\Delta\alpha_{n}=\sum_{i=1}^{N_{v}}\alpha_{n}^{(i)}\delta_{i,n}-\sum
_{j=N_{v}+1}^{N}\alpha_{n}^{(j)}\delta_{j,n}%
\]
Thus, matrix $\mathbf{J}_{m}(n)$ can be written as:

$\ $%
\begin{equation}
\mathbf{J}_{m}(n)=\left[
\begin{array}
[c]{ccccc}%
0 & 1 & 0 & \ldots & 0\\
0 & 0 & 1 & \ldots & 0\\
&  & \vdots &  & \\
0 & 0 & \ldots & 0 & 1\\
-\alpha_{n}^{tf}(\theta^{max}) & -\alpha_{n}^{tf}(\theta^{max-1}) & \ldots &
--\alpha_{n}^{tf}(\theta^{1}) & \Delta\alpha_{n}%
\end{array}
\right]  \label{matrix}%
\end{equation}
where $\theta^{max}=\max\{\theta^{(i)}\}$ and $\theta^{max-j}=\max
\{\theta^{(i)}\backslash\{\theta^{max},\ldots,\theta^{max-j+1}\}\}$. Equation
(\ref{Eq LAG}) finally becomes:
\begin{equation}
\mathbf{X}_{n+1}=\mathbf{J}_{m}\mathbf{\ }(n)\mathbf{X}_{n}+\mathbf{\eta}_{n}
\label{eq LAGmatrix final}%
\end{equation}
where
\begin{equation}
\mathbf{\eta}_{n}=\left[
\begin{array}
[c]{l}%
0\\
\vdots\\
-\eta_{n+1}+\alpha^{tf}\ \eta_{n}-\sum_{\theta=1}^{\theta_{max}}\alpha
_{n}^{tf}(\theta^{i})\ \eta_{n-\theta}%
\end{array}
\right]  \label{etatheta}%
\end{equation}
Equation (\ref{eq LAGmatrix final}), together with equation (\ref{matrix}) and
(\ref{etatheta}), define an Affine Random Dynamical System (see Apendix A and
\cite{Arnold98}).

Affine Random Dynamical Systems verify Oseledets's multiplicative ergodic
theorem \cite{Arnold98} and thus the Lyapunov numbers are the eigenvalues of
\[
\lim_{n\rightarrow\infty}\left[  \left(  J_{m}^{\ast}\right)  ^{n}\left(
J_{m}\right)  ^{n}\right]  ^{1/2n}%
\]
If the largest Lyapunov exponent of (\ref{eq LAGmatrix final}) is negative and
agents trade \emph{w.p.} $1$, then (\ref{eq LAGmatrix final}) is contractive
and $\mathbf{\eta}_{n}$ is a stationary process. This means that the
mispricing is a random walk and the returns follow a stationary process.

Notice that if $p^{(i)}\neq1$, $\mathbf{\eta}_{n}$ is not stationary and the
mispricing will not follow a random walk, as $\alpha_{n}^{tf}(\theta^{i})$ in
(\ref{etatheta}) will be time dependent. After the results of Section
\ref{sectionvi}, an interesting question is 'What is the asymptotic statistics
of $r_{n}$ if the traders' capital and activity probability are chosen so that
(\ref{market impact RC}) displays intermittent bursts?'.

For the returns' correlations to be small, trend followers and value investors
should have the same parameters \cite{Farmer98}. To simplify the analysis, we
will consider that these are trader independent. Thus each trader is
characterized by her strategy (value investor/trend follower), her capital,
$c$, and her activity probability, $p$. From equation (\ref{matrix}), we note
that $\left\langle \Delta\alpha_{n}\right\rangle =0$, so that the trend
followers' lag in time provides the oscillations observed on the system. As we
vary the traders' activity probability, $p$, equation (\ref{market impact RC})
approaches the critical point (zero Lyapunov exponent) equation (\ref{eq
LAGmatrix final}) displays on-off intermittent behaviour and the returns'
statistics shows two of the stylized facts of financial time series: fat tails
(indexes between $2$ and $4$) and volatility clustering.

The system was simulated for $N_{v}=N_{tf}=5\cdot10^{3}$, $\lambda=1$,
$\theta^{(i)}$ uniformly distributed between $\theta_{min}=1$ and
$\theta_{max}=100$, $\eta=0.1$, $c=8\cdot10^{-2}$ and $p=10^{-2}$. For these
parameters, the system is on the critical (non hyperbolic) point of zero
Lyapunov exponent and the statistics shown are highly non-trivial. The returns
distribution is fat tailed with an index of approximately $3$.

Nevertheless, it is difficult to infer from Fig. \ref{acf fig} whether the
decay of the autocorrelation function is exponential or power law. Let us
introduce the generalized correlations $C_{q}\left(  \tau\right)
$\nocite{Giulia2001}:
\[
C_{q}\left(  \tau\right)  \equiv\left\langle \left|  r_{t}\right|  ^{q}\left|
r_{t+\tau}\right|  ^{q}\right\rangle -\left\langle \left|  r_{t}\right|
^{q}\right\rangle \left\langle \left|  r_{t+\tau}\right|  ^{q}\right\rangle
\]
When the absolute returns series has long memory, $C_{q}\left(  \tau\right)  $
is a power law:%

\[
C_{q}\left(  \tau\right)  \sim\tau^{-\beta_{q}}%
\]
If $\left|  r_{t}\right|  ^{q}$ is an uncorrelated process one has $\beta
_{q}\sim1$, while $\beta_{q}$ less than $1$ corresponds to long range memory.
Let us introduce the \emph{generalized cumulative absolute returns }
\cite{Baviera2001}
\[
\chi_{t,q}\left(  \tau\right)  \equiv\frac{1}{\tau}\sum_{i=0}^{\tau-1}\left|
r_{t+\tau}\right|  ^{q}%
\]
and their variance
\[
\delta_{q}\left(  \tau\right)  \equiv\left\langle \chi_{t,q}\left(
\tau\right)  ^{2}\right\rangle -\left\langle \chi_{t,q}\left(  \tau\right)
\right\rangle ^{2}%
\]
Baviera et al. \cite{Baviera2001} show that if $C_{q}\left(  \tau\right)  $
for large $\tau$ is a power-law with exponent $\beta_{q}$, then $\delta
_{q}\left(  \tau\right)  $ is a power-law with the same exponent. That is
\[
C_{q}\left(  \tau\right)  \sim\tau^{-\beta_{q}}\Longrightarrow\delta
_{q}\left(  \tau\right)  \sim\tau^{-\beta_{q}}%
\]
In other words, the hypothesis of long range memory for the absolute returns
can be checked via the numerical analysis of the variance of the absolute
cumulative returns.

\begin{figure}
[ptb]
\begin{center}
\includegraphics[
height=3.8839in,
width=4.8525in
]%
{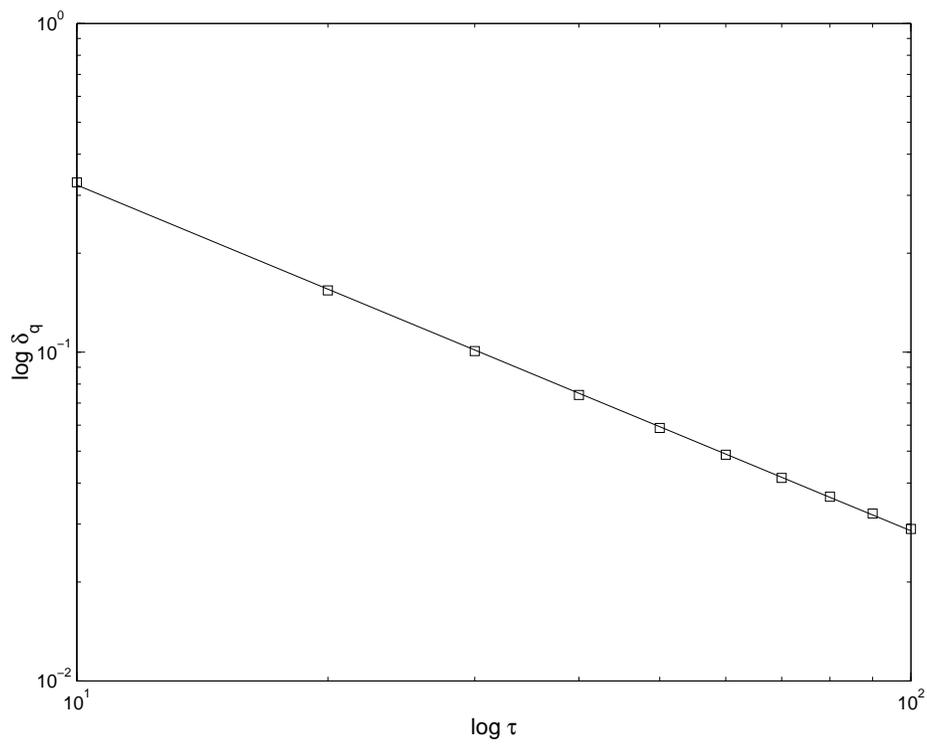}%
\caption{Volatility of cumulative absolute returns.}%
\label{fig corr}%
\end{center}
\end{figure}

In Fig \ref{fig corr} we plot the volatility of the absolute cumulative
returns. We determine an exponent $\beta_{1}\simeq1$, thus concluding that the
autocorrelation function has an exponential decay\footnote{Although hyperbolic
dynamical systems have exponentially decaying correlations, the reader should
note that our system has zero Lyapunov exponent, i.e. is non-hyperbolic.},
contrary to the stylized facts presented on page \pageref{intropage}. We
speculate that this is due to the intrinsic linearity of the model.

\section{Discussion}

\bigskip We have shown that when the dominant strategy is value investing, if
there exist $N,\alpha,$ $p$ and $\mu\in\left[  k,k+1\right]  $ that satisfy
(\ref{con alfa}) then
\begin{equation}
Prob\left(  \left|  r_{n}\right|  >x\right)  \approx c\cdot x^{-\mu}
\label{asymptotic}%
\end{equation}
and the returns display a probability distribution asymptotic to a power tail
of exponent $\mu$. We have also shown that, if one fixes $N$ and $p$, one can
determine heuristically $\alpha$ such that the probability distribution of
$r_{n}$ is given by (\ref{asymptotic}) with the desired $\mu$.

The association of an unconditional trading probability to each agent, permits
intermittent bursts on market price and a fat tailed distribution for the
returns. This is, to our knowledge, the first model that discusses fat tail
distributions caused by value investor strategies only.

We have shown that when trend followers and value investors are together on
the market trading at each time step, the mispricing is a random walk,
independently of the maximum lag in time, $\theta^{max}$ of the trend
followers. Trend followers do not seem to introduce relevant correlations on
the returns\footnote{If $\theta_{max}$ is large, the correlations induced by
trend followers on the returns compensate each other.}. A mixture of negative
short-term dependence (the value investors) with long-term positive dependence
(the trend followers) obscures the underlying dependence structure leading to
apparent insignificant correlations \cite{Farmer98}.

When the system is near the critical point (zero Lyapunov exponent) and each
trader has a relatively high capital and a low trading probability, the system
displays on-off intermittency. Further, the returns' distribution is fat
tailed and volatility is clustered.

We see our model as a first step towards an explanation of the stylized facts
of finance in the framework of a semi-analytical random dynamical system. The
feedback present in the model leads to volatility clustering, but is yet to be
improved so that the volatility autocorrelation function has a power-law decay.

Whether the present work is relevant and lessons for the market can be drawn
from it depends on the reliability of simulated price and return time series
(where it is noted that no mechanism provides any cointegration between price
and fundamental value) and on the acceptance of the underlying model
\cite{Farmer98}.

\section{Appendix A}

A random dynamical system (RDS) has two ingredients; $(i)$ a model of the
noise in the form of a measure-preserving transformation and $(ii)$ a model of
the deterministic dynamics consisting of a family of continuous
transformations. Which transformation is applied depends on the state of the
noise at that moment in time. We examine RDS generated by iterating skew
product maps of the form\cite{Ashwin99, Ashwin99a}%

\begin{align*}
\omega(t)  &  =\theta_{t}\omega\\
y\left(  t\right)   &  =\phi\left(  t,\omega\right)  y
\end{align*}
where $\theta_{t}$ is a flow and $\phi\left(  t,\omega\right)  $ is a cocycle
(i.e. is such that $\phi(t+s,\omega)=\phi(t,\theta_{s}\omega)\circ\phi\left(
s,\omega\right)  $ for all $t,s>0$). $\omega$ represents the state of a
dynamical system that models the noise process and $y$ represents the
dynamical system forced by the noise. We write $\theta_{t}\omega$ (resp.
$\phi\left(  t,\omega\right)  y$) to mean a nonlinear map $\theta_{t}$ (resp.
$\phi\left(  t,\omega\right)  $) applied to the point $\omega$ (resp. $y$). In
particular, the action of $\theta$ and $\phi$ need not be linear. In the case
that the evolution is chaotic we can see the above system as random forcing of
a deterministic system $\phi$ where the evolution of $\omega$ is `hidden'. By
looking at such systems one can get a more detailed picture of the dependence
of a dynamical system on noise than is possible by, for example, a
Fokker-Planck approach.

We assume that the evolution $\theta$ has an ergodic invariant probability
measure $\mathcal{P}$ with respect to which an initial condition for $\omega$
is chosen from a set of full measure. Random dynamical systems studies the
dynamics of the full system relative to the dynamics of the measure-preserving
transformation $(\theta,P)$ ( $\mu$ is defined on subsets in some $\sigma$-algebra).

A random dynamical system is said to be affine if
\[
\phi\left(  t,\omega\right)  y=\Phi\left(  t,\omega\right)  y+\psi\left(
t,\omega\right)
\]
where $\Phi\left(  t,\omega\right)  $ is a linear cocycle and
\[
\psi\left(  t+s,\omega\right)  =\Phi\left(  t,\theta_{s}\omega\right)
\psi\left(  s,\omega\right)  +\psi\left(  t,\theta_{s}\omega\right)  ,\qquad
t,s\geq0
\]

\section{Appendix B}

\noindent\textsl{Proof of Lemma \ref{lemma0}: }The \emph{k}th moment of
$\alpha_{n}^{\mu}$ is given by
\begin{equation}
M^{(k)}(0)=\sum_{i=0}^{N}i^{k}p^{i}(1-p)^{N-i}\left(
\begin{array}
[c]{l}%
N\\
i
\end{array}
\right)  \label{Moment k}%
\end{equation}
A Taylor series expansion of (\ref{Moment k}) in $p$ to second order about the
origin yields\footnote{We observe that only terms $i<3$ contribute to a second
order approximation in $p$.}
\[
M^{(k)}(0)=\left\{
\begin{array}
[c]{ll}%
\qquad Np & k=1\\
Np\left[  1+(2^{k-1}-1)(N-1)p\right]  +O^{3}(p) & k\geq2
\end{array}
\right.
\]
So that
\begin{equation}%
\begin{array}
[c]{lll}%
\left(  \frac{M^{(k+1)}(0)}{M^{(k)}(0)}\right)  ^{k} & = & \left(
\frac{1+(2^{k}-1)(N-1)p}{1+(2^{k-1}-1)(N-1)p}\right)  ^{k}\\
& = & \left(  1+\frac{2^{k-1}(N-1)p}{1+(2^{k-1}-1)(N-1)p}\right)  ^{k}%
\end{array}
\label{Quocient Moment}%
\end{equation}
We want to prove that for $p$ small enough
\[
\left(  \frac{M^{(k+1)}(0)}{M^{(k)}(0)}\right)  ^{k}>M^{(k)}(0)
\]
Which is equivalent to
\begin{equation}
\left(  1+\frac{2^{k-1}(N-1)p}{1+(2^{k-1}-1)(N-1)p}\right)  ^{k}>\left(
1+(2^{k-1}-1)(N-1)p\right)  Np \label{Inequality Moments}%
\end{equation}
Expanding both sides in a Taylor series in $p$ about the origin, we obtain the
relation
\[
1+2^{k-1}k(N-1)p+O^{2}(p)>Np+O^{2}(p)
\]
which, to first order in $p$, is equivalent to
\[
N>\frac{2^{k-1}k}{2^{k-1}k-1}-\frac{1}{\left(  2^{k-1}k-1\right)  p}%
\]
which is valid for $k>1$ and $N>1$.$\blacksquare$

\noindent\textsl{Proof of Lemma \ref{lemma1}: }

The solution of (\ref{eq vi}) is given by \cite{Haan89}
\[
r_{n}=\sum_{k=1}^{n}\left(  \prod_{j=k+1}^{n}-\alpha_{j}\right)  \alpha
_{k}\eta_{k}+\left(  \prod_{j=1}^{n}-\alpha_{j}\right)  r_{0}%
\]
where $\prod_{j=n+1}^{n}-\alpha_{j}=1$. The Lyapunov number of (\ref{eq vi})
is given by
\begin{equation}
\left\langle \alpha_{n}\right\rangle =\alpha M^{(1)}(0)=\alpha Np
\label{LyapExponent}%
\end{equation}
If $\left\langle \alpha_{n}\right\rangle <1$ (negative Lyapunov exponent)
\[
\prod_{j=1}^{n}-\alpha_{j}^{vi}\underset{n\rightarrow\infty}{\rightarrow}0
\]
exponentially fast and under very weak conditions on the product $\alpha
_{k}\eta_{k}$, the distribution of $r_{n}$ will converge independently of
$r_{0}$ to that of the series
\begin{equation}
\sum_{k=1}^{n}\left(  \prod_{j=k+1}^{n}-\alpha_{j}\right)  \alpha_{k}\eta_{k}
\label{limiting distribution}%
\end{equation}
Thus, if $\alpha<\frac{1}{M^{(1)}(0)}$, $r_{n}$ converges in distribution to a
unique limiting distribution, that of (\ref{limiting distribution}). As the
binomial moments $M^{(k)}(0)$ are increasing functions of $k$ this condition
is satisfied for $\alpha\in I_{k,k+1}\quad\forall k\in\mathbb{N}
$.$\mathbb{\blacksquare}$

As can be observed from (\ref{limiting distribution}), the additive term in
(\ref{eq vi}), $\alpha_{n}\eta_{n}$, provides a reinjection mechanism,
allowing $r_{n}$ to fluctuate without converging to zero, as it would if
$\alpha_{n}\eta_{n}$ vanished \cite{Sornette97}.

\noindent\textsl{Proof of Proposition \ref{proposition}}\emph{: }If the
agents' normalized capitals and trading probabilities are constants,
respectively $\alpha_{n}^{(i)}=\alpha$ and $p^{(i)}=p$,
\begin{equation}
\left\langle \left(  \alpha_{n}^{vi}\right)  ^{\mu}\right\rangle =\alpha^{\mu
}\sum_{i=0}^{N}i^{\mu}\,\,p^{i}(1-p)^{N-i}\binom{N}{i} \label{moments}%
\end{equation}
Theorem \ref{theorem Kesten} requires the existence of a $\mu>0$ such that
\begin{equation}
\left\langle \left(  \alpha_{n}^{vi}\right)  ^{\mu}\right\rangle =1
\label{momentu}%
\end{equation}
Although the estimation of $\mu$ from (\ref{momentu}) is in general not
possible, one can estimate an interval for $\mu$. Choose $p\in\left]
0,1\right[  $ and $\{N,k\}\in\mathbb{N}^{2}$. As the binomial moments
$M^{(k)}(0)$ are continuous and increasing functions of $k$, equation
(\ref{momentu}) is satisfied if there exists an $\alpha$ such that
\begin{equation}
\left\{
\begin{array}
[c]{c}%
\left\langle \left(  \alpha_{n}^{vi}\right)  ^{k}\right\rangle =\alpha
^{k}M^{(k)}(0)<1\\
\left\langle \left(  \alpha_{n}^{vi}\right)  ^{k+1}\right\rangle \alpha
^{k+1}M^{(k+1)}(0)>1
\end{array}
\right.  \label{con alfa}%
\end{equation}
From (\ref{moments}) and (\ref{con alfa}), one deduces that $\alpha\in
I_{k,k+1}$.$\blacksquare$

\section{\medskip Appendix C}

Suppose $\left\{  X_{n},n\geq1\right\}  $ is a stationary sequence and that
\[
P\left[  X_{1}>x\right]  =x^{-\alpha}L\left(  x\right)  ,\qquad x\rightarrow
\infty
\]
where $L$ is slowly varying and $\alpha>0$. Let
\[
X_{\left(  1\right)  }>X_{\left(  2\right)  }>\cdots>X_{\left(  n\right)  }%
\]
be the order statistics of the sample $X_{1},\cdots,X_{n}$. We pick $k<n$ and
define the Hill estimator to be \cite{Adler98}
\[
H_{k,n}=\frac{1}{k}\sum_{i=1}^{k}\log\frac{X_{\left(  i\right)  }}{X_{\left(
k+1\right)  }}%
\]
where $k$ is the number of upper order statistics used in the estimation. The
Hill plot is the plot of
\[
\left(  \left(  k,H_{k,n}^{-1}\right)  ,1\leq k<n\right)
\]
and, if the processs is linear or satistifies mixing conditions, then since
$H_{k,n}\overset{P}{\rightarrow}\alpha^{-1}$ as $n\rightarrow\infty$,
$k/n\rightarrow0$ the Hill plot should have a stable regime sitting at height
roughly $\alpha$.

As an alternative to the Hill plot, it is sometimes useful to display the
information provided by the Hill estimation as \cite{Resnick99}
\[
\left\{  \left(  \theta,H_{\left\lceil n^{\theta}\right\rceil ,n}^{-1}\right)
,0\leq\theta\leq1\right\}  ,
\]
where we write $\left\lceil y\right\rceil $ for the smallest integer greater
or equal to $y\geq0$. We call such plots the \emph{alternative Hill plot}. The
alternative display is sometimes revealing since the original order statistics
get shown more clearly and cover a bigger portion of the displayed space.

\bibliographystyle{THESNUMB}
\bibliography{bib}
\newpage%

\begin{figure}
[ptb]
\begin{center}
\includegraphics[
height=3.9539in,
width=4.9095in
]%
{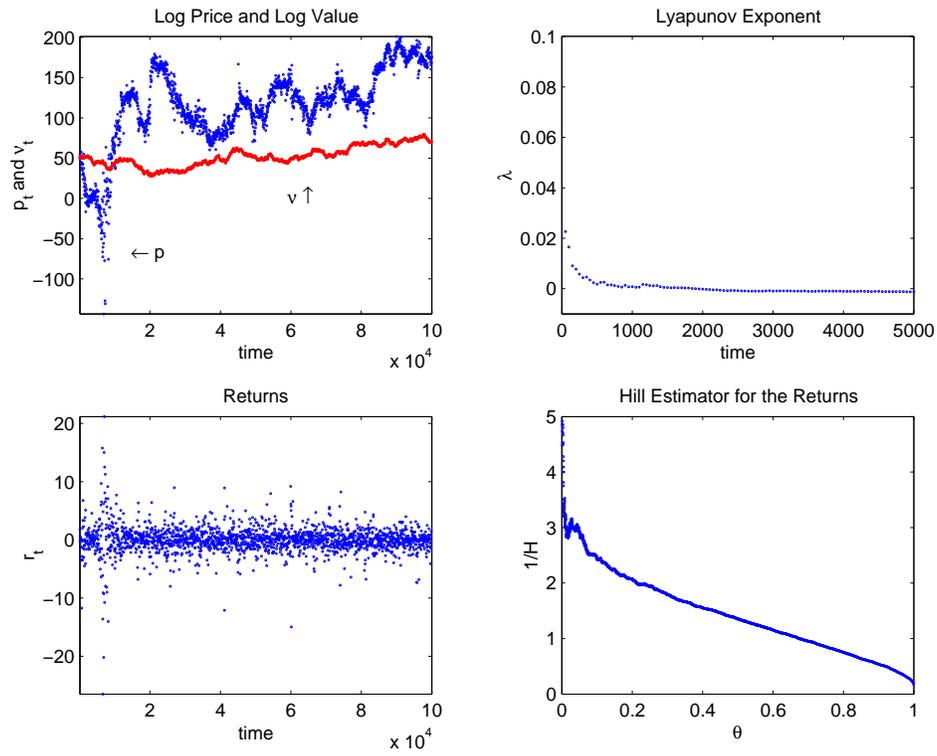}%
\caption{Simulation with $N_{v}=N_{tf}=5\cdot10^{3}$, $\lambda=1$, $\eta=0.1$,
$\theta_{min}=1$ and $\theta_{max}=100$, $c=8\cdot10^{-2}$ and $p=10^{-2}$.}%
\label{composite fig}%
\end{center}
\end{figure}

\begin{figure}
[ptb]
\begin{center}
\includegraphics[
height=3.8467in,
width=4.8724in
]%
{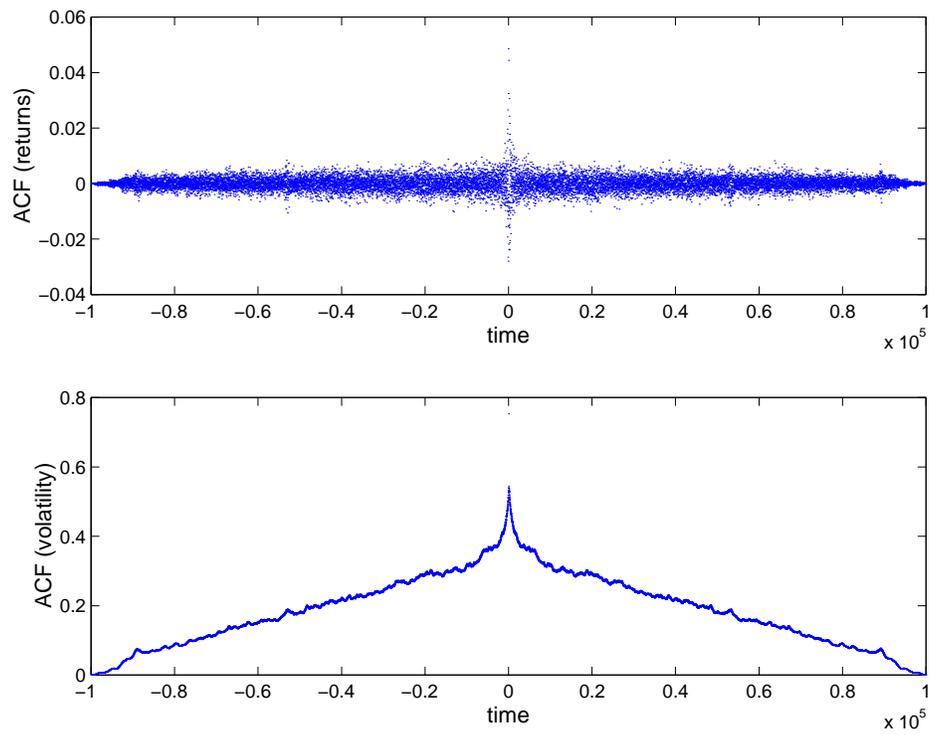}%
\caption{Return and Volatility Autocorrelation Functions for the case in Fig
\ref{composite fig}.}%
\label{acf fig}%
\end{center}
\end{figure}

\end{document}